\documentclass[12pt,preprint]{aastex}
\usepackage{graphicx}
\newcommand{\Halpha}{H$\alpha$}
\newcommand{\kms}{km\,s$^{-1}$}

\begin{document}
\slugcomment{Accepted for publication: Astronomische Nachrichten,
  (AN), 2013, 334, p. 73 ({\it Cambridge Workshop 17 on Cool Stars, Stellar
  Systems and the Sun})}

\title{New insights: the accretion process and variable wind \\ from TW
 Hya\footnote{Data were obtained with the CHANDRA satellite, the 6.5m Magellan/Clay
        telescope at Las Campanas Observatory, and Gemini-S which is
 operated by the Association of Universities for Research in Astronomy, Inc. under
 a cooperative agreement with the US-NSF on behalf of the Gemini partnership.}}

\author{A. K. Dupree}
\affil{Harvard-Smithsonian Center for Astrophysics, 60 Garden
  Street, Cambridge, MA 02138 USA}
\email{dupree@cfa.harvard.edu}

\begin{abstract}
For the first time in a classical T Tauri star, we are able to trace an accretion event signaled 
by an hour-long enhancement of X-rays from the accretion shock and 
revealed through substantial sequential  changes in optical emission line profiles. 
Downflowing turbulent material appears in H$\alpha$ and H$\beta$
emission.  He D3 (5876\AA) broadens, coupled with an increase in flux.   
Two hours after the X-ray accretion event, the 
optical veiling increases due to continuum emission from the hot 
splashdown region.  The response of the stellar coronal emission to 
the heated photosphere follows about 2.4 hours later, giving direct 
evidence that the stellar corona is heated in part by accretion. 
Then, the stellar wind becomes re-established.  
A model that incorporates the dynamics of this sequential series of 
events includes: an accretion shock, a cooling downflow in a 
supersonically turbulent region, followed by photospheric 
and later, coronal heating. This model naturally explains the presence 
of broad optical and ultraviolet lines, and affects the mass accretion 
rates currently determined from emission line profiles.  
These results, coupled with the large heated coronal region revealed from 
X-ray diagnostics, suggest that  current models are not adequate to 
explain the accretion process in young stars. 
  \end{abstract}

\keywords{stars: individual (TW Hya) -- stars: mass-loss -- stars:
  variables: T Tauri stars -- stars: winds, outflows -- X-rays: stars}

\section{Introduction}

TW Hya (CD $-$34 7151; HIP 53911) is arguably the closest accreting T
Tauri star making it a choice object for detailed spectroscopic study.  
It is generally thought that cool material from the circumstellar disk
surrounding the star is channeled by the stellar magnetic field
and free-falls towards the star where a shock forms.  The X-ray and
optical spectra
reported here probe this accretion process. Another
major attribute of TW Hya is the fact that its surrounding
circumstellar  disk appears roughly face-on (Qi et al. 2004) so that the critical polar
regions, where the accretion process occurs, can be observed directly without
obscuration by the disk.  

A major observational campaign was aimed towards TW Hya that
involved dedicated X-ray spectroscopic observations amounting 
to 500 total ks with the CHANDRA spacecraft.  These results are
discussed in Brickhouse et al. (2010).   Contemporaneous optical and infrared spectroscopy
and photometry were carried out from 4 continents (Dupree et al. 2012). 
As a result, the process of accretion can be investigated, and for the 
first time, the source of the broad emission 
lines from the star can be reliably identified.

\section{X-ray accretion signatures}

The CHANDRA X-ray  spectra contain many emission lines arising from high
temperatures. Analysis of the line strengths reveals that three 
different components producing X-rays  must be  present in the TW Hya 
atmosphere: a high temperature ($\sim$10$^7$ K) corona,
a lower temperature (3 $\times$ 10$^6$K) component arising from 
the accretion shock, and a large volume producing \ion{O}{7}, at slightly lower
temperature (2.5$\times$10$^6$K)  with density lower than
the accretion shock  itself (Brickhouse et al. 2010).   The behavior
of lines arising in the accretion shock, namely \ion{N}{7}, \ion{O}{8}, \ion{Ne}{9}, \ion{Fe}{17},
and \ion{Mg}{11}, can give a direct measure of the instantaneous strength of the
accreting material.  These lines are marked in the CHANDRA spectrum
shown in Fig 1, and the variation of the strength of the sum of the
accretion lines is shown in Fig. 2

The accretion line flux (Fig. 2) reveals an enhancement 
centered at JD 2454157.75 that
represented the highest  count rate in a  3 ks average
of the long 500ks CHANDRA observation, and merits special study.

\section{Simultaneous optical spectra}

Many simultaneous high-resolution optical
spectra were taken during the X-ray accretion enhancement, providing the opportunity
to evaluate the effect of the increase in the accreting line flux on 
the optical emission lines.  Echelle spectra were taken
continuously over 3 nights with the MIKE spectrograph at the Magellan/CLAY telescope 
of Las Campanas Observatory. These spectra have a 
resolution $\sim$35,000 so that the line profiles are
well-resolved.  Fig. 3 shows the behavior
of the total flux and the line profile asymmetries during the
span of the X-ray measures shown in Fig. 2.  The flux
of H$\alpha$ does not exhibit any systematic change,
which is not surprising since the line is
surely optically thick.  However, the asymmetry of the
line does  change quite abruptly following
the X-ray accretion event.  The asymmetry of
a line profile indicates the mass flow in a 
differentially moving line-forming region (Hummer \& Rybicki 1968).
Of course, if a slab moves at constant velocity, the whole
profile will shift by an amount corresponding to the constant
velocity. But  differential motion as found  in  a stellar
wind or in downflowing material  produces a change in the line asymmetry.
If the short-wavelength side (``blue'') of the
line is stronger than the long-wavelength side (``red''), material
is flowing away from the observer and {\it vice versa}. 
The abrupt increase in the 
value of the blue:red ratio for H$\alpha$ suggests an abrupt
increase in down-flowing material that begins about 9 minutes
after the increase in the
X-ray accretion line flux.  This increased
inflow continues for about 1.5 hours.  The H$\beta$ emission
line exhibits a similar increase in the ratio of
blue to red emission, echoing the increased downflow
of material exhibited by the H$\alpha$ line.  In addition,
an increase in the strength of the total H$\beta$ emission
occurs.

The D3 line of \ion{He}{1} at 5876\AA\ is also a valuable probe of the accretion
process.  It is known to have both a broad and narrow component of
emission. The broad component is generally thought to signal
accretion (Donati et al. 2011) in a similar way as  the broad lines of the
Balmer series.  A sharp increase in the flux at the end of the X-ray
accretion event results from an enhanced long wavelength wing of
the line profile, and the increase in line flux beginning about
30 minutes later (see Fig. 4) is due to a 30\% increase in the broad component
of the line, whereas the narrow component is constant to within
15\%. The sequential changes of the optical lines following the X-ray
accretion strongly suggests that they form in the post-shock cooling 
zone.

The line widths themselves give additional support to this formation
scenario.  The Balmer lines have a FWHM of about $\pm$150 \kms, which 
is obviously in excess of a thermal width ($\sim$21 \kms) at a 
temperature of 10$^4$K, and also in harmony with the measured
widths of the \ion{Ne}{9} ($\pm$165 \kms) lines observed in the CHANDRA
spectra (Brickhouse et al. 2010).   Far ultraviolet line
widths of \ion{C}{3}  and \ion{O}{6}, when corrected for wind absorption,
suggest similar line widths of $\pm$160 \kms\ (Dupree et al. 2005).

The veiling or `weakening' of absorption lines in the optical spectrum 
arises from a continuum and perhaps a contribution from line emission
(Petrov et al. 2011)
produced by the  accretion hot spot in the photosphere.  The value of
the veiling from the short wavelength region (4400--5000\AA) of the MIKE spectra 
is shown in Fig 4 also.  It too increases with a delay of 2 hours
after
the X-ray accretion event.  Such a delay is consistent with a
reasonable
size of the photospheric `hot spot'. Using a value of 35 \kms\ for
the post shock downflow indicated by substantial absorption at that
velocity in the H$\beta$ profile suggests that material will traverse
a distance comparable to the size of a hot spot covering $\sim$10\% of
a star with radius 0.8$R_\odot$ in 2.8 hours.  The flux from the
corona (represented by the first order CHANDRA spectrum) responds
$\sim$2 hours later to an increase in veiling (Dupree et al. 2012).

\section{The stellar wind}
The Balmer profiles also reveal the wind structure and its variation. 
\Halpha\ and H$\beta$ 
profiles observed over 4 successive nights are shown in Fig. 5 and
Fig. 6.  The X-ray accretion event discussed 
earlier occurred during the first night when  the \Halpha\ profile is roughly 
symmetric. During the subsequent 3 nights, absorption
appears and systematically increases on the negative velocity side of
the line. The wind which appears very weak or perhaps absent on the first
night, recovers and becomes more opaque during the following nights.

The symmetry of the broad \Halpha\ profile on the first night demonstrates
that the line is not formed in an accretion stream (which should
appear only at positive velocities),  but most likely  in a 
turbulent region of the atmosphere with velocity centered on the
TW Hya itself.  \Halpha\ has a higher opacity than H$\beta$ and 
would be formed higher in the atmosphere (`at the edges') of a 
turbulent region than the H$\beta$ transition.  Thus it is not
surprising that the inflow signature is stronger in the weaker
H$\beta$ line.

Because the wind also substantially modifies the emission line profiles, 
this suggests  that the
line width may not be a good indicator of the accretion 
rate as has been proposed (Natta et al. 2004).  In fact,
in our observations, the wind (and undoubtedly accretion contributes 
also) changes the width of the line at the 10\% level which, if 
dependent only on accretion, corresponds to
a factor of 5 in the mass accretion rate.

The near-infrared \ion{He}{1} transition at $\lambda$10830 has proved to
be an excellent tracer of winds from T Tauri stars (Dupree et
al. 2005).  This transition arises from a metastable level of
neutral helium and thus maintains a relatively high population which
can absorb the strong infrared continuum from a cool star.  The
broad emission appears likely to arise also in the post-shock cooling
zone of TW Hya, and the extent of the absorption (to $\sim -$300 \kms)
clearly documents the presence of  a fast wind.  The terminal velocity 
of the wind varies with time   ($-$260 \kms\ in 1992, Dupree et 
al. 2005; $-$330 \kms\ in 2002, Edwards et al. 2006; $-$300 \kms\ in 2007).  Lines formed at
even higher kinetic temperatures (\ion{O}{6} and \ion{C}{3}) also give evidence
of similar high outflow velocities (Dupree et al. 2005)  suggesting that this hot wind 
 may be powered as a result of the accretion 
process  which acts as a source of energy and 
momentum in the upper atmosphere (Matt et al. 2012; Cranmer 2009).

\section{A new model}

These observations reveal a distinct new view of the accretion process
in classical T Tauri stars and suggest that the current 
models need revision.
It has been common to attempt modeling
of the emission features in the spectrum as arising from
the accretion stream that is channeled by the
magnetic field as it approaches free-fall velocity
forming an accretion shock (Muzerolle et al. 2000; Kurosawa et
al. 2011).  The  spectra shown here suggest
that an accretion event (observed in X-rays) instigated
a cascade of changes to the optical emission line profiles.
The emission lines arise in the post-shock cooling
volume, and the line widths from X-ray profiles (Brickhouse et al. 2010), 
far UV (Ardila et al. 2002), the optical
and the near-IR helium line shown here are all commensurate and broad.  Their
breadth has been difficult to interpret in the framework of
accreting streams (Ardila et al. 2002) or an accretion shock 
(Lamzin et al. 2007).  A turbulent post-shock cooling zone 
offers the likely solution to the puzzle of the broad profiles, and
is supported by the behavior of the optical lines.  The emission
measure of the post-shock cooling zone exceeds that of the accretion
stream by a factor of 100 (Dupree et al. 2012).   
Another component of the post-shock cooling process is the discovery
from the \ion{O}{7} diagnostics in CHANDRA spectra  (Brickhouse et al. 2010)
of a large coronal region with 300 times the volume and 30 times the emission
measure of the accretion shock.  All of these observations call for
a reassessment of current models of accretion in young stars.  It
appears that the accretion process can both heat the corona, cause turbulent
broadening of the emission lines, and provide a means to power 
an accretion-driven stellar wind.

\clearpage


\begin{figure}
\begin{center}
\includegraphics[angle=0, scale=0.7]{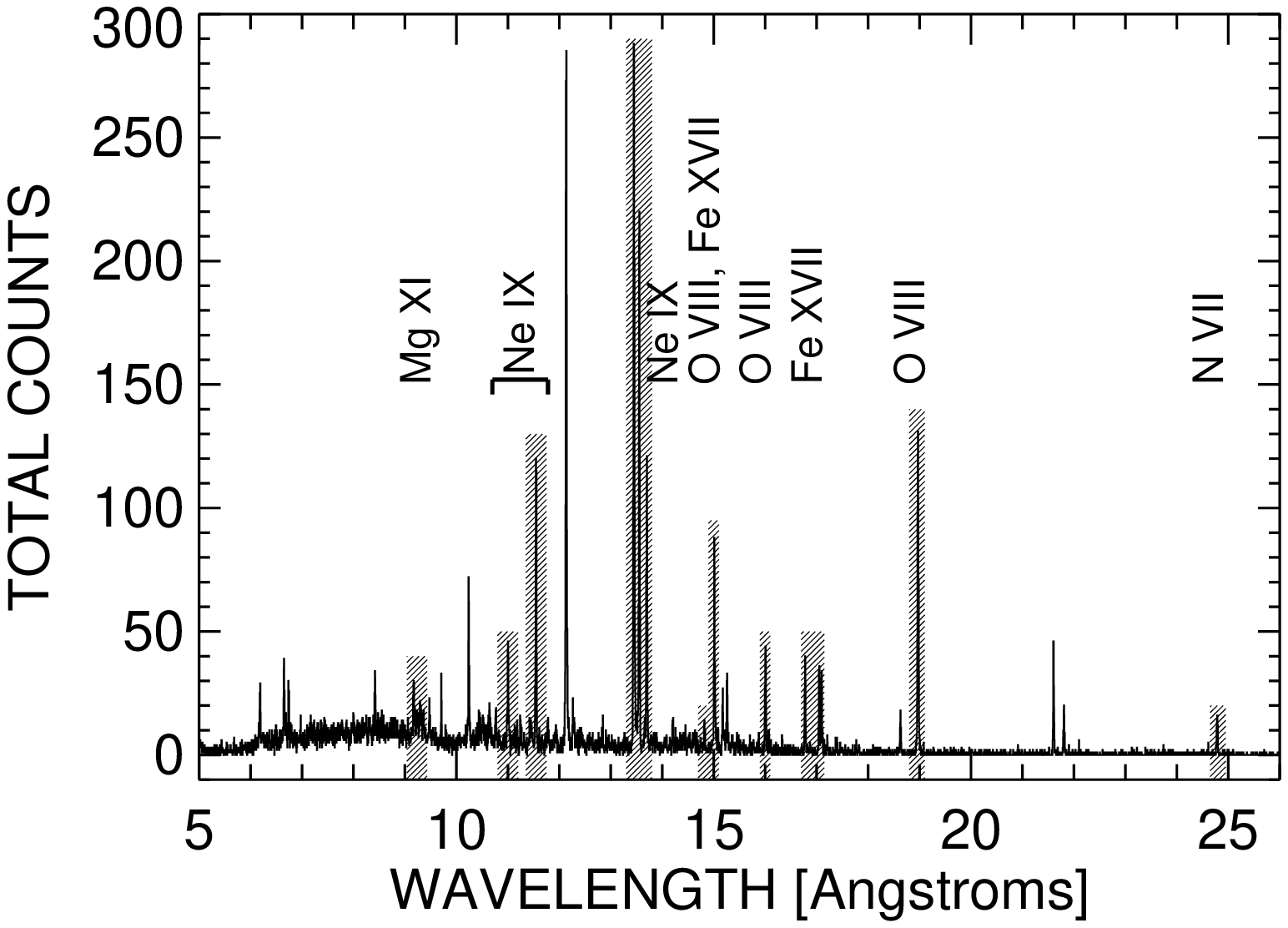}

\caption{Lines arising in the accretion shock (\ion{N}{7}, \ion{O}{8}, \ion{Ne}{9},
  \ion{Fe}{17}, \ion{Mg}{11}) marked in the MEG total spectrum (hatched areas).}

\end{center}
\end{figure}

\begin{figure}
\includegraphics[angle=90, width=\linewidth]{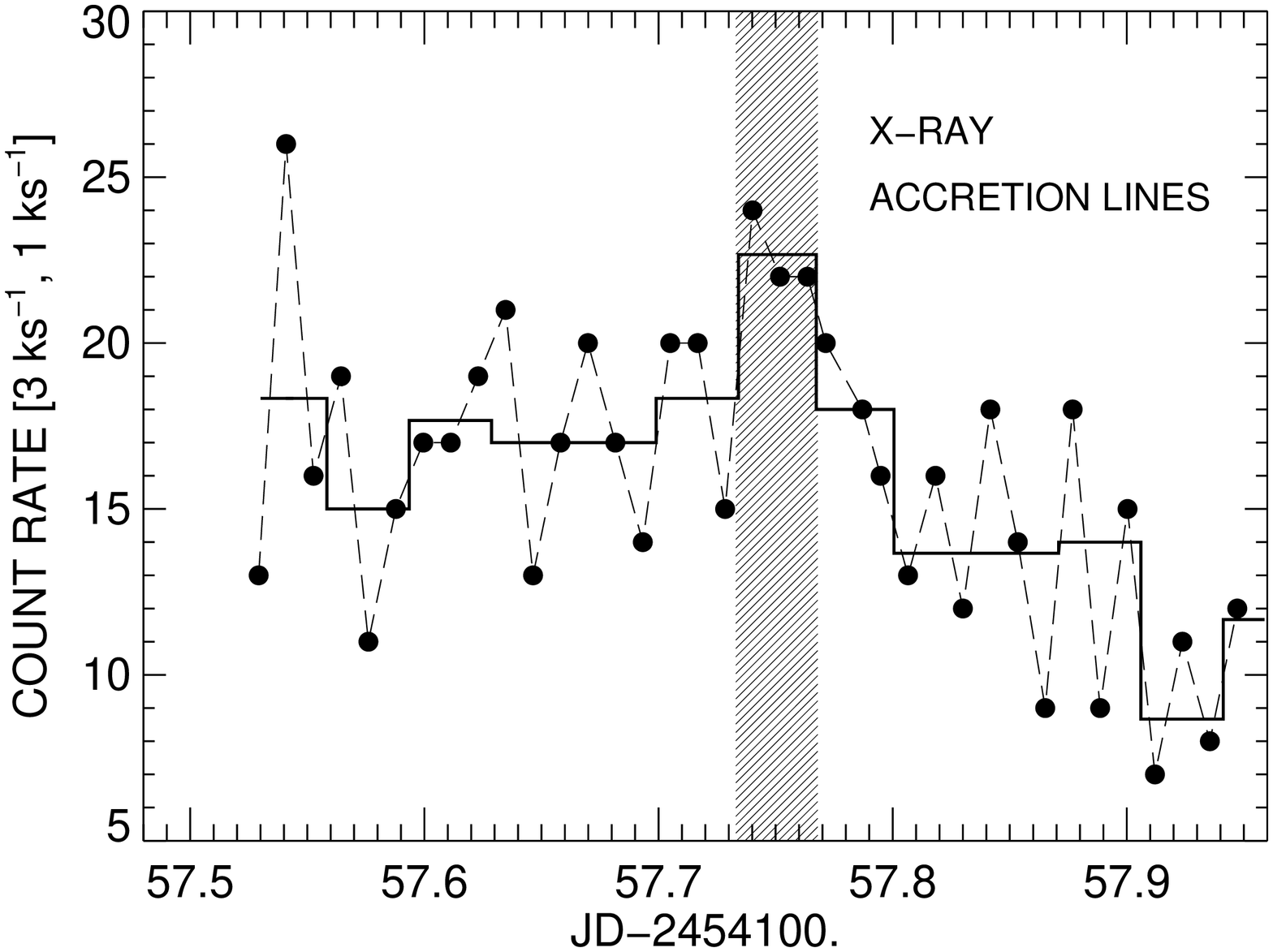}

\caption{Accretion lines (\ion{N}{7}, \ion{O}{8}, \ion{Ne}{9},
  \ion{Fe}{17}, \ion{Mg}{11}) binned over 1 ks (filled circles)
  or 3 ks and divided by 3 for display (solid line).  The
hatched region marks the accretion event and is carried forward 
in the following figures.}
\label{Flabel2}
\end{figure}


\begin{figure}
\includegraphics[angle=0, width=\linewidth]{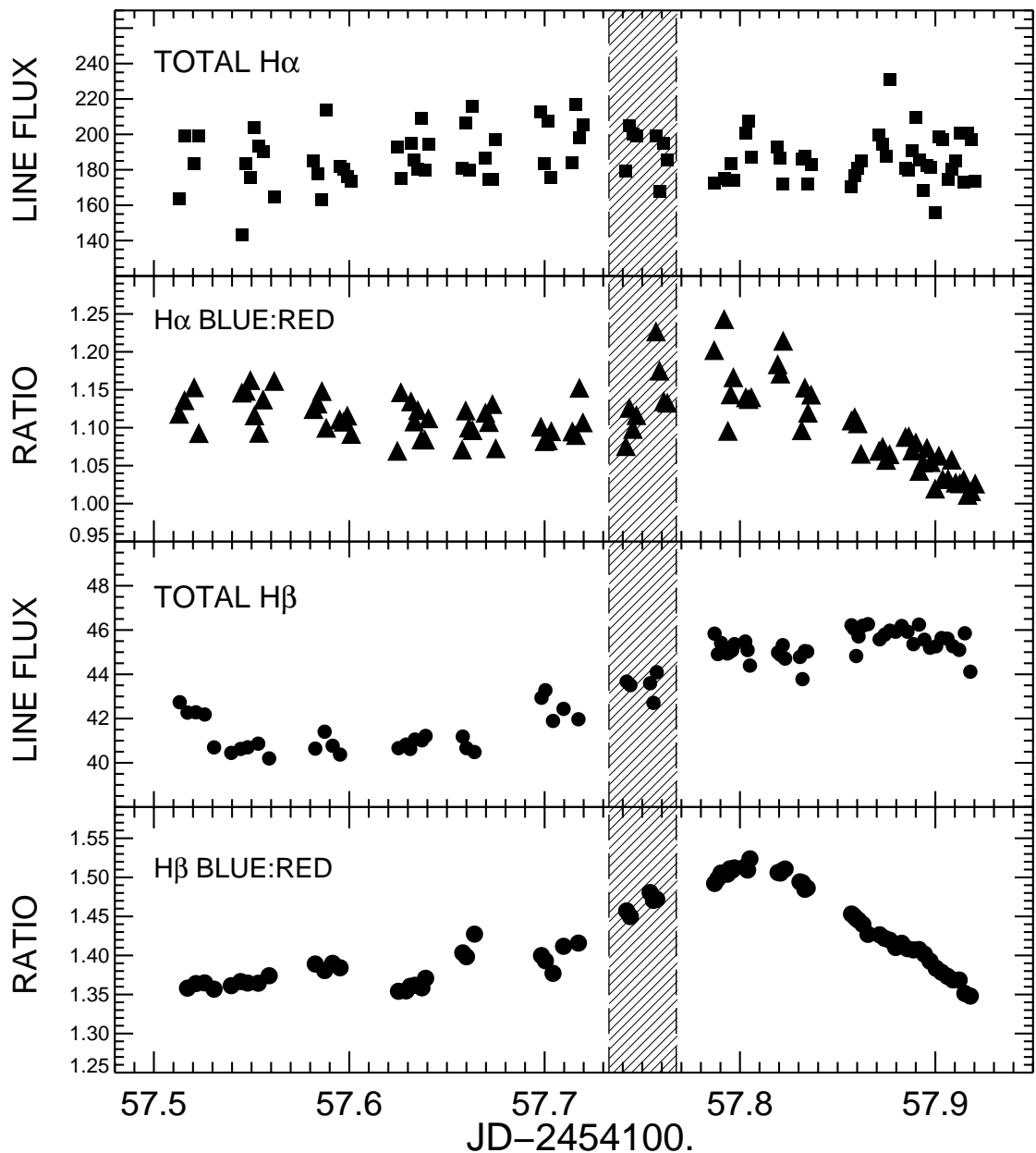}

\caption{The behavior of the H$\alpha$ and H$\beta$ line fluxes and
 profiles.  The X-ray accretion event  is marked by
 line hatching.  The ratio  ``blue:red'' denotes the flux in  1\AA\
 bandpasss positioned $-$2\AA\ (``blue'') or $+$2\AA\ (``red'')  from line center in
 continuum-normalized spectra.
}
\label{Flabel3}
\end{figure}


\begin{figure}
\includegraphics[angle=0, width=\linewidth]{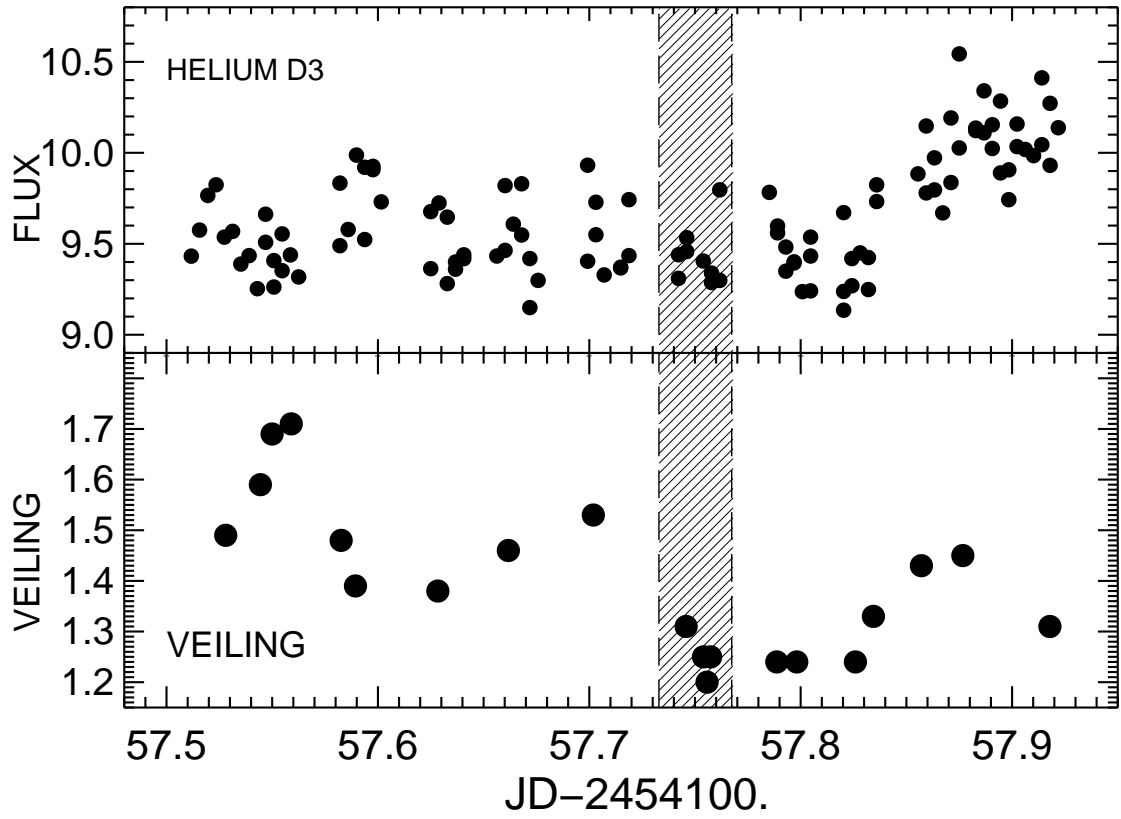}

\caption{The behavior of the  \ion{He}{1} 5876\AA\ (D3) flux and the value of the average
blue veiling derived from the MIKE spectra.}
 
\label{Flabel4}
\end{figure}

\begin{figure}
\includegraphics[angle=90, width=\linewidth]{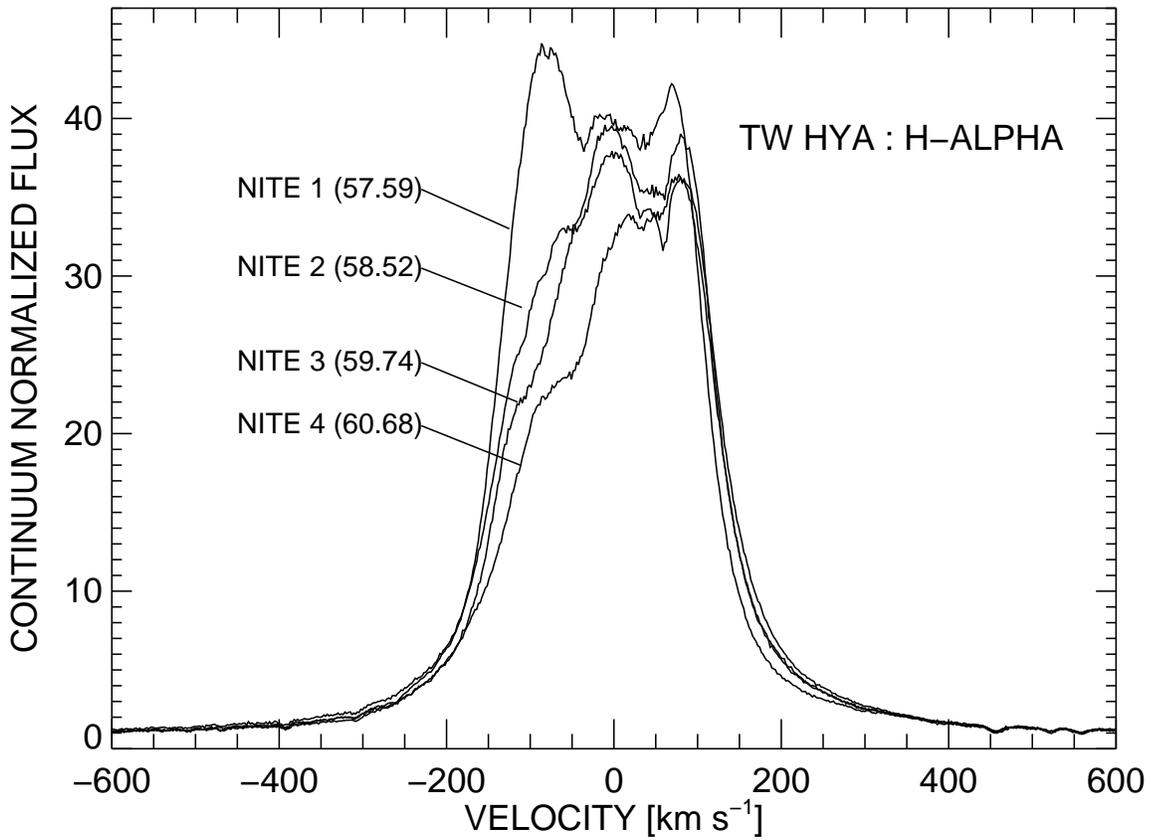}

\caption{The H$\alpha$ profile over 4 successive nights.  Note the
  systematic apparent onset and strengthening of the wind absorption
  on the negative velocity side of the profile. The Julian date of
  each spectrum is marked as JD$-$2454100.  These profiles were taken
from the red side spectra of MIKE.}
 
\label{Flabel5}
\end{figure}

\begin{figure}
\includegraphics[angle=90, width=\linewidth]{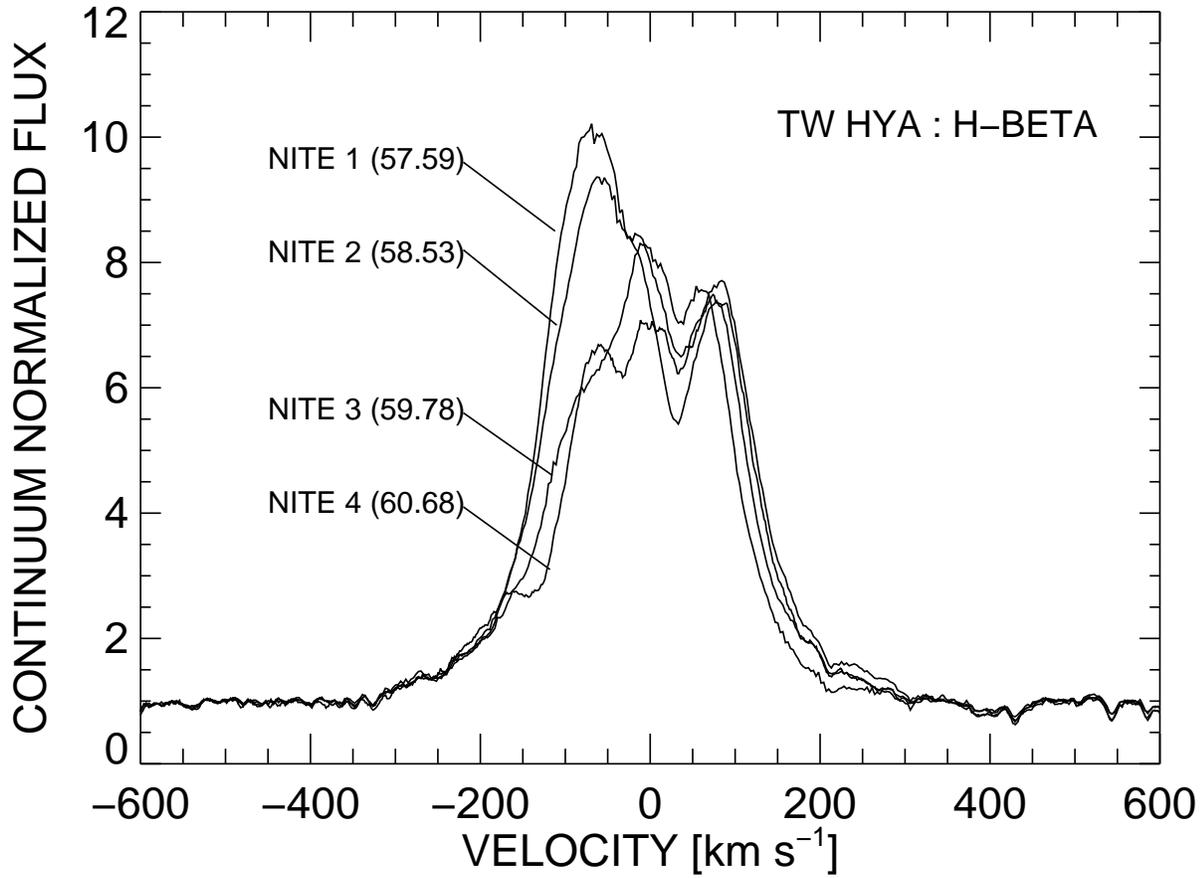}

\caption{The H$\beta$ profile over 4 successive nights.  Note the
similarity to \Halpha\ with the onset and strengthening of
wind absorption.  H$\beta$ exhibits a clear infall signature with 
absorption at a velocity of $\sim$ +35 \kms. These profiles were taken from the
blue side spectra of MIKE. The Julian date of
  each spectrum is marked as JD$-$2454100.}
\label{Flabel6}
\end{figure}

\begin{figure}
\begin{center}
\includegraphics[angle=0, scale=0.8]{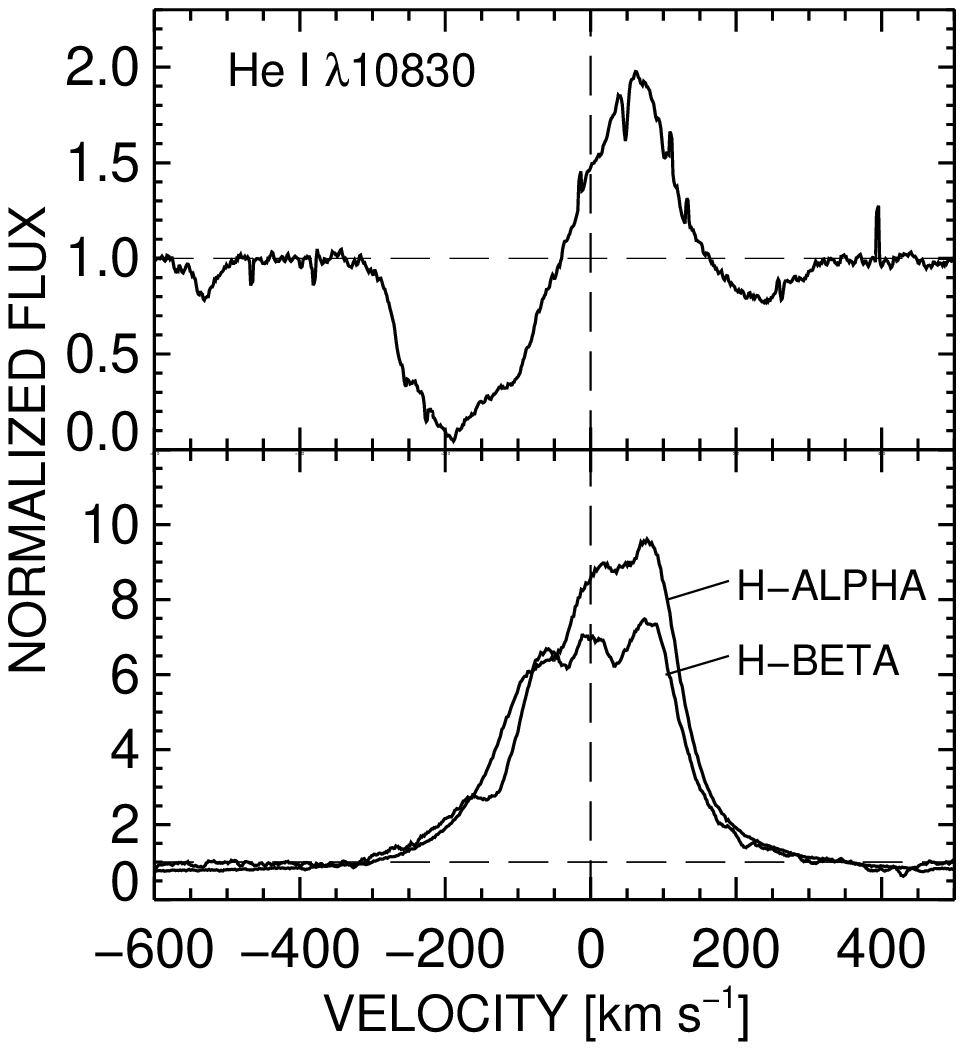}
\caption{The wind-sensitive \ion{He}{1} $\lambda$10830 line taken
with PHOENIX on Gemini-S on the same night (JD 2454160.7) and within an hour of the H$\alpha$
and H$\beta$ spectra (Dupree et al. 2012).  The near-IR helium
absorption shows that the wind extends to $\sim$ $-$300~\kms, and
the emission is weaker than found in previous years.  In addition
stronger downflow (extending to +300~\kms) than found  previously 
(Dupree et al. 2005) is indicated 
by  absorption on the positive velocity side of the profile. At the
same time, such high velocities are not visible in the Balmer series,
although the wind clearly modifies these lines at lower velocities
(0 to $-$200 \kms).   The H$\alpha$ profile is reduced for display 
in this figure by a factor of 4.}
\end{center}
\label{Flabel7}

\end{figure}


\newpage

\begin{thebibliography}{}

\bibitem[Ardila (2002)]{ar02}
Ardila, D. R., Basri, G., Walter, F. M., Valenti, J. A., Johns-Krull,
C. M.  2002, ApJ, 566, 1100

\bibitem[Brickhouse (2010)]{br10}
Brickhouse, N. S., Cranmer, S. R., Dupree, A. K., Luna, G. J. M.,
Wolk, S.  2010, ApJ, 710, 1835 

\bibitem[Cranmer (2009)]{cr09}
Cranmer, S. R.  2009, ApJ, 706, 824


\bibitem[Donati (2011)]{don11}
Donati, J.-F., Gregory, S. G., Alencar, S. H. P., et al.  2011, MNRAS, 417, 472

\bibitem[Dupree (2012)]{dup12}
Dupree, A. K., Brickhouse, N. S., Cranmer, S. R., et al.  2012, ApJ, 750, 73 



\bibitem[Dupree (2005)]{dup05}
Dupree, A. K., Brickhouse, N. S., Smith, G. H., Strader, J.  2005,
ApJ, 625, L131


\bibitem[Edwards (2006)]{ed06}
Edwards, S., Fischer, W., Hillenbrand, L., Kwan, J.  2006, ApJ, 646, 319

\bibitem[Hummer (1968)]{hu68}
Hummer, D. G., Rybicki, G. B.  1968, ApJ, 153, L107

\bibitem[Kurosawa (2011)]{ku11}
Kurosawa, R., Romanova, M. M., Harries, T. J.  2011, MNRAS, 416, 2623

\bibitem[Lamzin (2007)]{la07}
Lamzin, S. A., Romanova, M. M., Kravtsova, A. S.  2007, in  J. Bouvier,
I. Appenzeller (eds.), {\it  Star-Disk Interaction in Young Stars}, IAU
Symp. 243 , p. 115


\bibitem[Matt (2012)]{ma12}
Matt, S. P., Pinz\'on, G., Greene, T. P., Pudritz, R. E.  2012, ApJ,
745, 101

\bibitem[Muzerolle (2000)]{muz00}
Muzerolle, J., Calvet, N., Brice\~no, C., Hartmann, L., Hillenbrand,
L.  2000, ApJ, 535, L47

\bibitem[Natta (2004)]{nat04}
Natta, A., Testi, L., Muzerolle, J., Randich, S., Comer\'on, F.,
Persi, P.  2004, A\&A, 424, 603

\bibitem[Petrov (2011)]{pe11}
Petrov, P. P., Gahm, G. F., Stempels, H. C., Walter, F. M., Artemenko,
S. A.  2011, A\&A, 535, A6

\bibitem[Qi (2004)]{qi04}
Qi, C., Ho, P., Wilner D., et al  2004, ApJ, 616, L11

\end{thebibliography}
\end{document}